\documentclass[structabstract]{aa}  
\pdfoutput=1
\usepackage{graphicx}
\usepackage{txfonts}
\usepackage{natbib}
\bibpunct{(}{)}{,}{a}{}{,}
\usepackage{subfigure}

\begin{document}
   \title{New Parameters and Transit Timing Studies for OGLE2-TR-L9 b}

   \author{M. Lendl
	  \inst{1,3,5}
	  \and
	  C. Afonso
	  \inst{1}
	  \and
	  J. Koppenhoefer
	  \inst{2,6}
	  \and
	  N. Nikolov
	  \inst{1}
	  \and
	  Th. Henning
	  \inst{1}
	  \and
	  M. Swain
	  \inst{4}
	  \and
          J. Greiner
	  \inst{6}
}

   \institute{Max Planck Institute for Astronomy, 
              K\"onigstuhl 17, 69117 Heidelberg, Germany
	      \and
	      Universit\"ats-Sternwarte M\"unchen, Scheinerstr. 1, 81679 Munich, Germany
	      \and
	      Institut f\"ur Astronomie, Universit\"at Wien, T\"urkenschanzstr. 17, 1180 Wien, Austria
	      \and
	      Jet Propulsion Laboratory, California Institute of Technology, 4800 Oak Grove Drive, Pasadena, California 91109-8099, USA
	      \and
	      Observatoire de Gen\`eve, Universit\'e de Gen\`eve, 51 chemin des Maillettes, 1290 Versoix, Switzerland
	      \and
	      Max Planck Institute for Extraterrestrial Physics, Giessenbachstr., 85748 Garching, Germany
}

   \date{}

 
  \abstract
{Repeated observations of exoplanet transits allow us to refine the planetary parameters and probe them for any time dependent variations.
In particular deviations of the period from a strictly linear ephemeris, transit timing variations (TTVs), can indicate the 
presence of additional bodies in the planetary system.}
   {Our goal was to reexamine the largely unstudied OGLE2-TR-L9 system with 
high cadence, multi-color photometry in order to refine the planetary parameters and probe the system for TTVs.}
   {We observed five full transits of OGLE2-TR-L9 with the GROND instrument at the ESO/MPG 2.2 m telescope at La Silla Observatory. 
GROND is a multichannel imager that allowed us to gather simultaneous light curves in the g', r', i', and z' filters.
    }
   {From our analysis we find that the semi-major axis and the inclination differ from the previously published values. 
With the newly observed transits, we were able to refine the ephemeris to $2454492.80008(\pm0.00014) + 2.48553417(\pm6.4 \times 10^{-7})E$.
The newly derived parameters are $a=0.0418\pm0.0015$ AU, $r_{p}=1.67\pm0.05$ $R_{j}$, and $inc=82.47^\circ\pm0.12$, differing 
significantly in $a$ and $inc$ from the previously published values. Within our data, we find indications for TTVs. }
   {}


   \maketitle
%

\section{Introduction}
In the study of extrasolar planets, the observation of planetary transits has a prominent position. 
It allows us to determine several parameters that are not accessible by other means, shedding more light 
on the nature of the transiting planet. From the transit light curve itself, the planetary radius and 
the orbital inclination can be found by investigating its shape and depth. 
Together with radial velocity measurements, the knowledge of the inclination allows us to determine 
the true mass of the planet, which can then be used in conjunction with the radius to find the planet's 
mean density, allowing its composition to be constrained.
With spectroscopic observations during transit and secondary eclipse, the chemical composition of
 planetary atmospheres can be studied \citep[e.g.][]{Charbonneau02,Swain08} 
. 

Despite its advantages and increasing success, the transit method presents limitations for small and farther 
out planets: it is strongly biased towards close-in planets since the geometrical probability of a planet to transit 
decreases with orbital separation and the transit signal becomes much less distinct for small planets, reaching 
down to a transit depth of only $0.01\%$ for an Earth-sized planet orbiting a solar-type star. 
With increasing orbital periods, scheduling becomes a major constraint on the discovery of planets, as transit signals
become sparser in time, so long-term observations are required.

The transit timing method has the potential to detect small transiting or non transiting objects 
such as moons, Trojans, or additional planets that can be difficult to find
with current methods. It makes use of the fact that the gravitational impact of these objects can be measured as dynamical interactions 
within the system, causing the time between successive transits to vary, and in turn the transit occurs earlier or later than 
expected \citep{Holman05}. The transit timing method aims at discovering additional objects by searching for changes in the 
mid-transit times of close-in transiting exoplanets. Naturally, this technique is more sensitive to massive perturbers but 
it can be sensitive enough to detect planets of only a few Earth masses which are located near mean motion resonances with the transiting giant. 
Several searches for TTVs in exoplanetary systems have been undertaken up to now \citep[e.g.][]{Alonso09,Diaz08,Holman06}.
While no additional objects have been found with this technique, it has been used to pose limits on the existence of such objects in several
cases \citep[e.g.][]{Rabus09,Adams10}.
To detect TTVs one needs to observe a large number of transits and build up a database of mid-transit 
times covering several seasons. Additionally, the repeated observation of transits enables us to refine the planetary parameters.

The subject of this paper, OGLE2-TR-L9 b, has been identified as a planetary candidate in the publicly available 
data of the OGLE-II project \citep{Udalski97}
by \citet{Snellen07}. It was recently confirmed as a transiting exoplanet with the observation of
a full transit and radial velocity measurements by \citet{Snellen09}. 
OGLE2-TR-L9 b has a mass of $4.5\pm1.5 M_{J}$ and a period of approximately 2.5 days \citep{Snellen09} orbiting
 an F3V star, at this point in time the hottest star known to host a transiting planet.

In this paper, we present five new transits of OGLE2-TR-L9 observed with the GROND instrument \citep{GROND} mounted at the ESO/MPG
2.2 m telescope at La Silla Observatory during April and May 2009. We recorded each transit simultaneously in four optical channels, g', r', i', and z'.
Using this newly available data and reanalyzing the data obtained by \citet{Snellen09}, we redefine the planetary parameters and conduct a first search for TTVs. An overview of the observations and the data reduction methods used will be given in Section \ref{sec:obsdata}, while a description of 
the data analysis and error estimation is presented in Section \ref{sec:detsys}. Section \ref{sec:res} summarizes the obtained results.


\section{Observations and Data Reduction} 
\label{sec:obsdata}
\subsection{Observations}

We observed five transits of OGLE2-TR-L9 using the GROND (\textbf{G}amma \textbf{R}ay Burst \textbf{O}ptical and 
\textbf{N}ear-Infrared \textbf{D}etector) instrument \citep{GROND} mounted at the 
ESO/MPG 2.2 m telescope at La Silla Observatory. Designed with the aim of observing gamma-ray-burst 
afterglows, GROND simultaneously observes in 4 optical (\emph{g' r' i' z'}) and 3 infrared (\emph{J H K}) channels 
allowing us to gather 4 optical light curves of each transit event. In the infrared channels, series of exposures of 
10 sec each were obtained. Since dithering was switched off for a large part of the observations to improve time sampling and 
photometric precision, no stacking of the infrared frames is possible with the standard way of background subtraction. We therefore
ignore the \emph{JHK} channels in our analysis. 

The observations took place during the nights of
April 10, 15, 20, 25 and May 15 2009, corresponding to the Epochs 177, 179, 181, 183 and 191 based on the ephemeris given by
\citet{Snellen09}. During each night, we observed the full transit plus at least 20 min of baseline before
ingress and after egress. For all observations, the exposure time was kept fixed at \mbox{46.4 s}. During the night 
of April 10, we used the slow \mbox{($\sim45$ s)} readout mode achieving a cadence of \mbox{2.05 min}. We incurred some problems 
with guiding during the night of April 15, leading to unstable data quality. During this night, we were able 
to test the fast \mbox{($\sim15$ s)} readout mode and observed with a cadence of \mbox{1.95 min.}
For the nights of April 20, 25 and May 15, we used the fast readout mode, increasing
the cadence to \mbox{1.31 min} for April 20 and \mbox{1.25 min} for April 25 and May 15.  

In our analysis, we included the transit observed with GROND on January 27, 2008 by \citet{Snellen09}. The observation procedure was 
essentially the same as in our first observation (April 10). A detailed description of the observations can be found in 
\citet{Snellen09}, while the data reduction was redone for this work as described below.

\subsection{Data Reduction}

With one of the goals of the initial OGLE observations being the detection of micro lensing events, OGLE2-TR-L9 is located in a 
rather crowded field. Fortunately, OGLE2-TR-L9 is reasonably bright (I=13.97 mag) and fairly isolated, with only very minor sources in 
its vicinity allowing the use of aperture photometry. 

The image calibration was performed with the \emph{mupipe} software which has been developed at the University 
Observatory Munich\footnote[1]{\emph{mupipe} is available from http://www.usm.uni-muenchen.de/people/arri/mupipe/},
and encompassed overscan and flatfield corrections. Then we carried out flux measurements for 
aperture photometry using IRAF\footnote[2]{IRAF is distributed by the National Optical Astronomy Observatories,
which are operated by the Association of Universities for Research
in Astronomy, Inc., under cooperative agreement with the National Science Foundation.} daophot procedures. 
We iteratively selected approximately 15 of the most stable stars in the field (with the exact number depending on filter and 
seeing conditions) and combined them to a reference source to be used for differential photometry.

\begin{figure*}[ht]
\centering
\subfigure[]{
\includegraphics[scale=0.5]{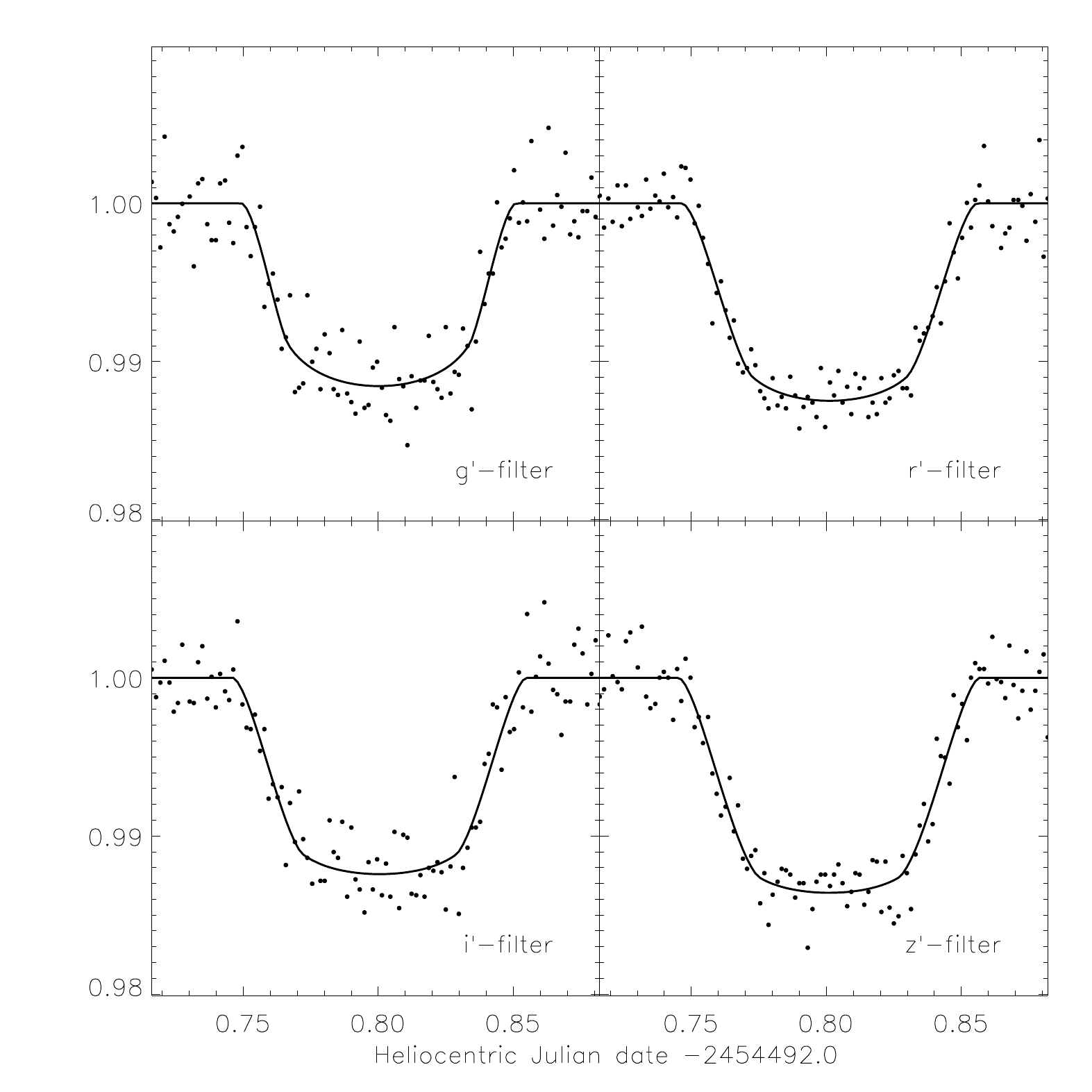}
\label{fig:subfigO}
}
\subfigure[]{
\includegraphics[scale=0.5]{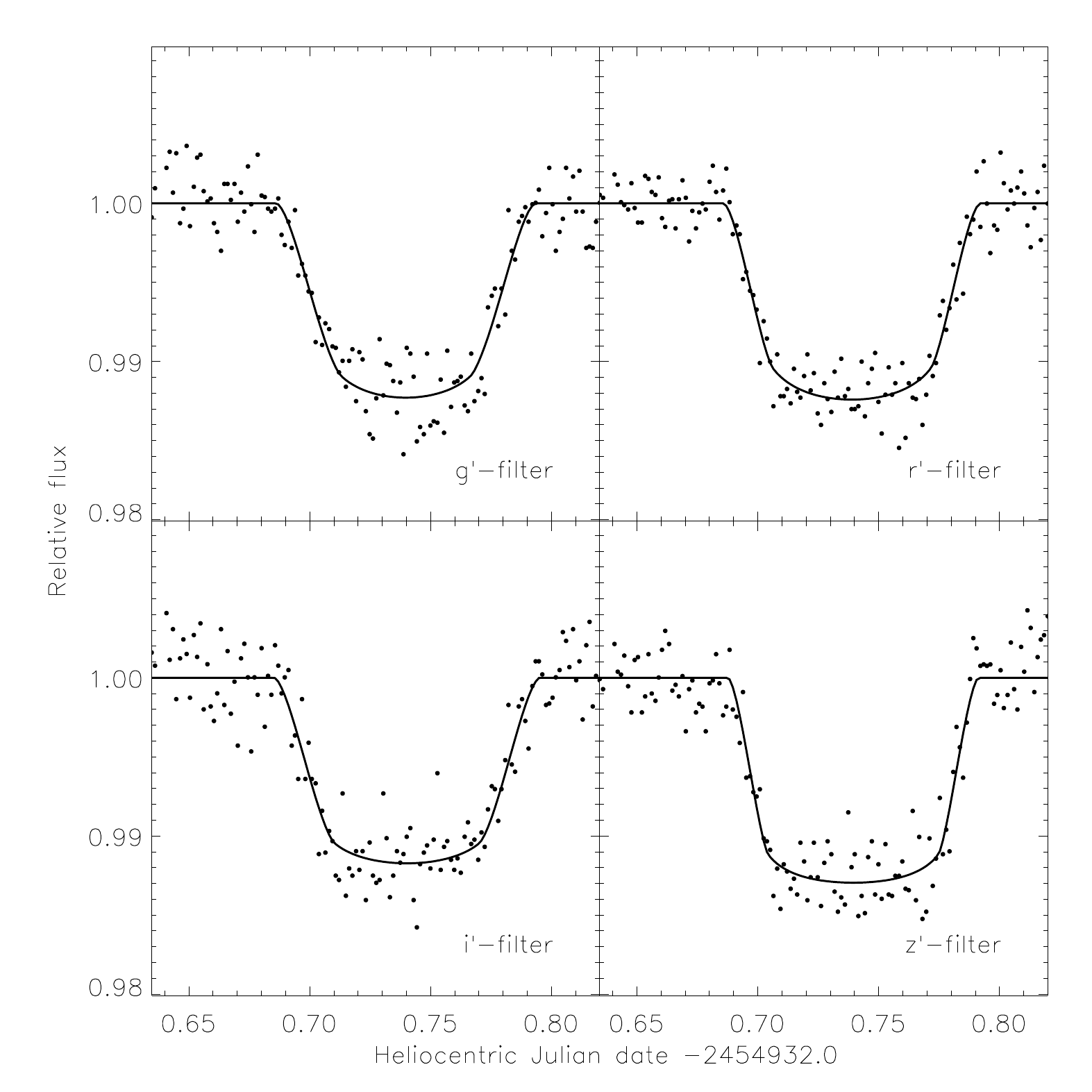}
\label{fig:subfig1}
}

\subfigure[]{
\includegraphics[scale=0.5]{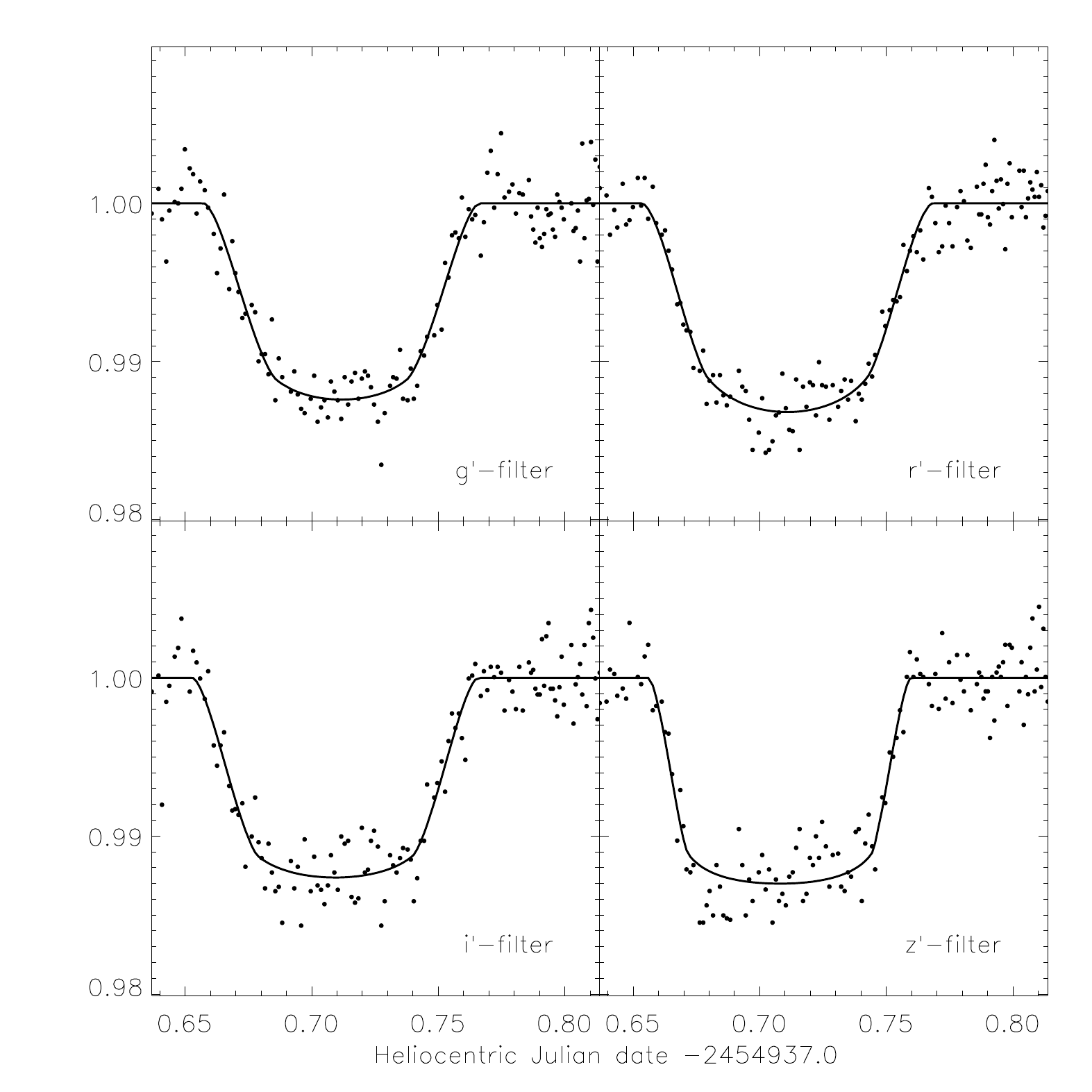}
\label{fig:subfig2}
}
\subfigure[]{
\includegraphics[scale=0.5]{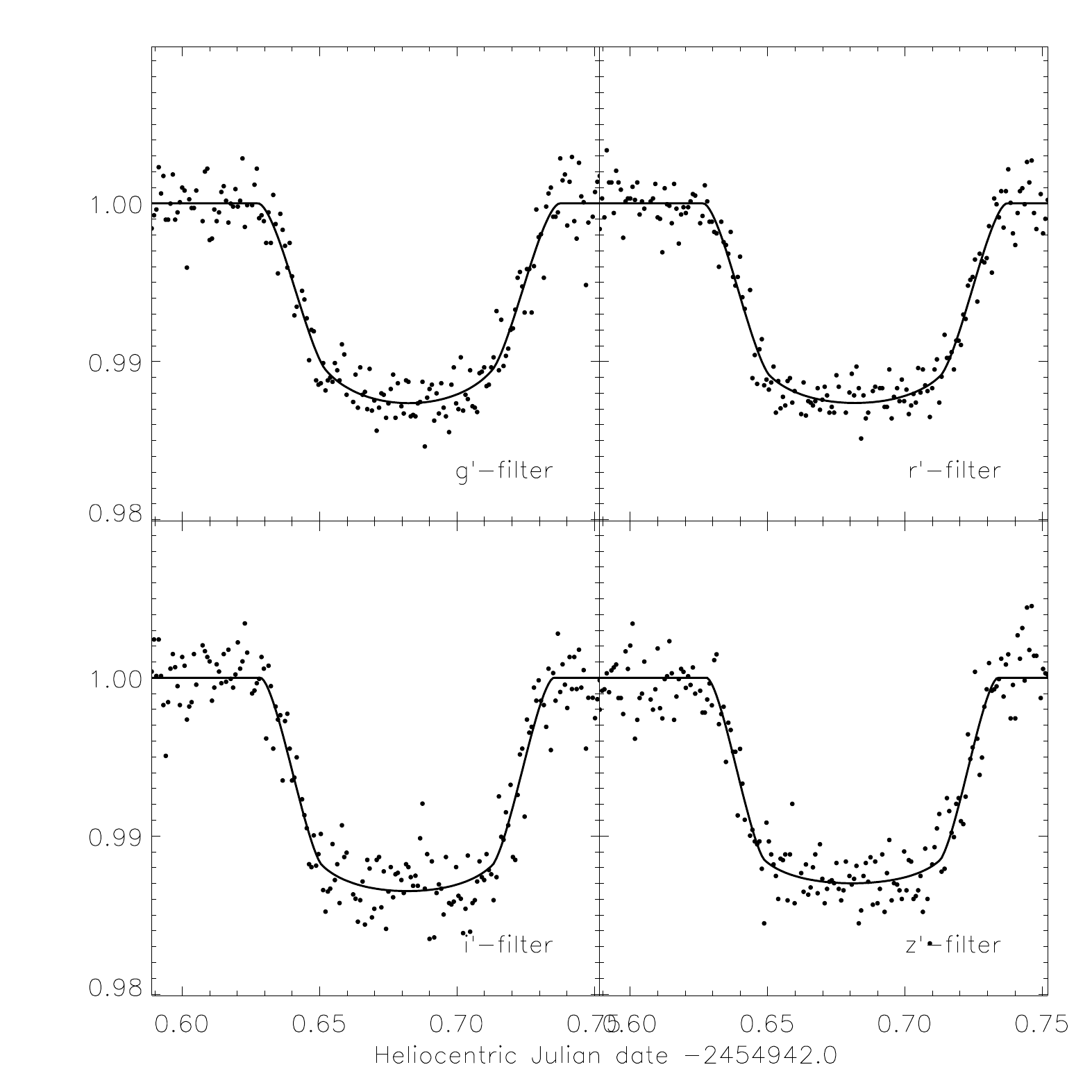}
\label{fig:subfig3}
}

\subfigure[]{
\includegraphics[scale=0.5]{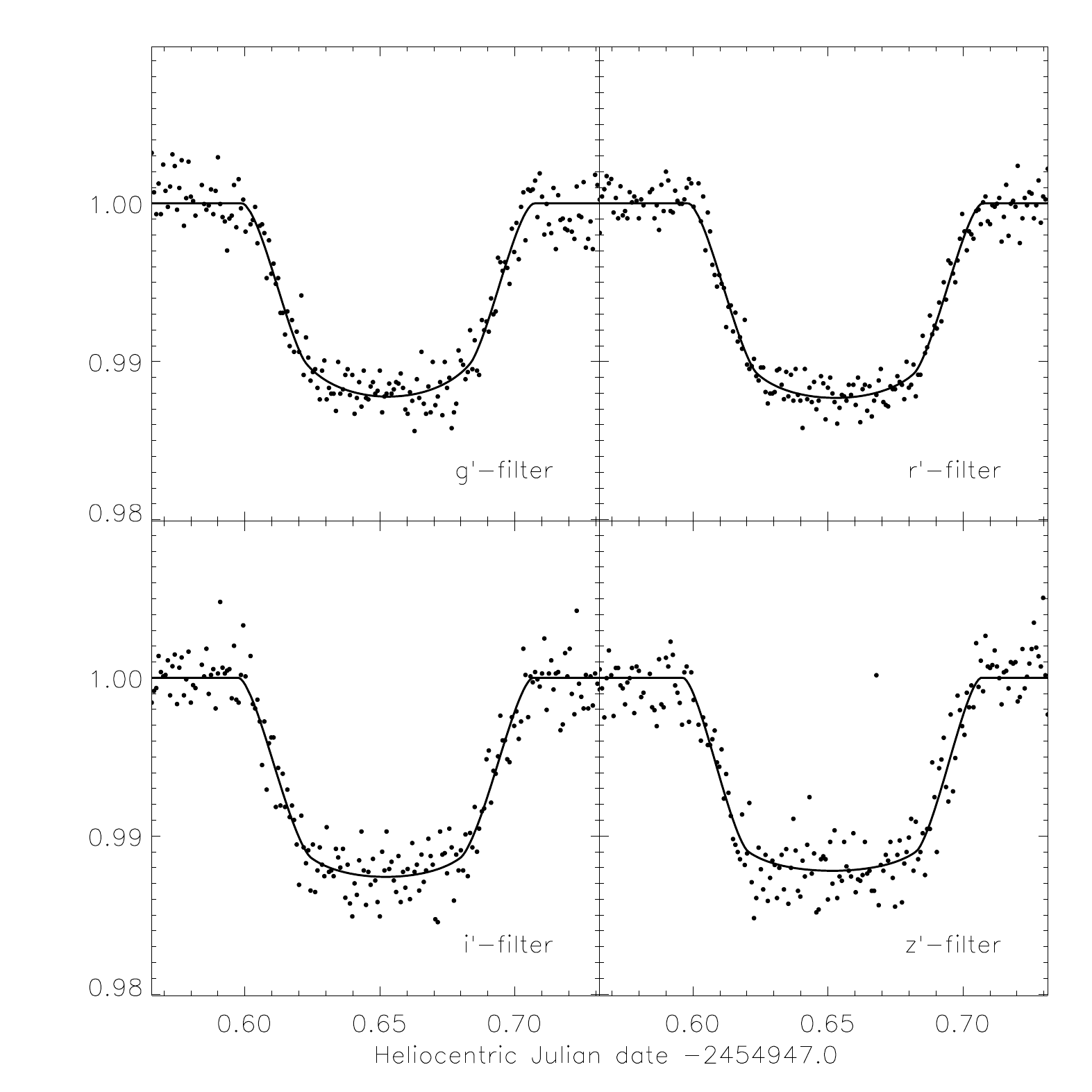}
\label{fig:subfig4}
}
\subfigure[]{
\includegraphics[scale=0.5]{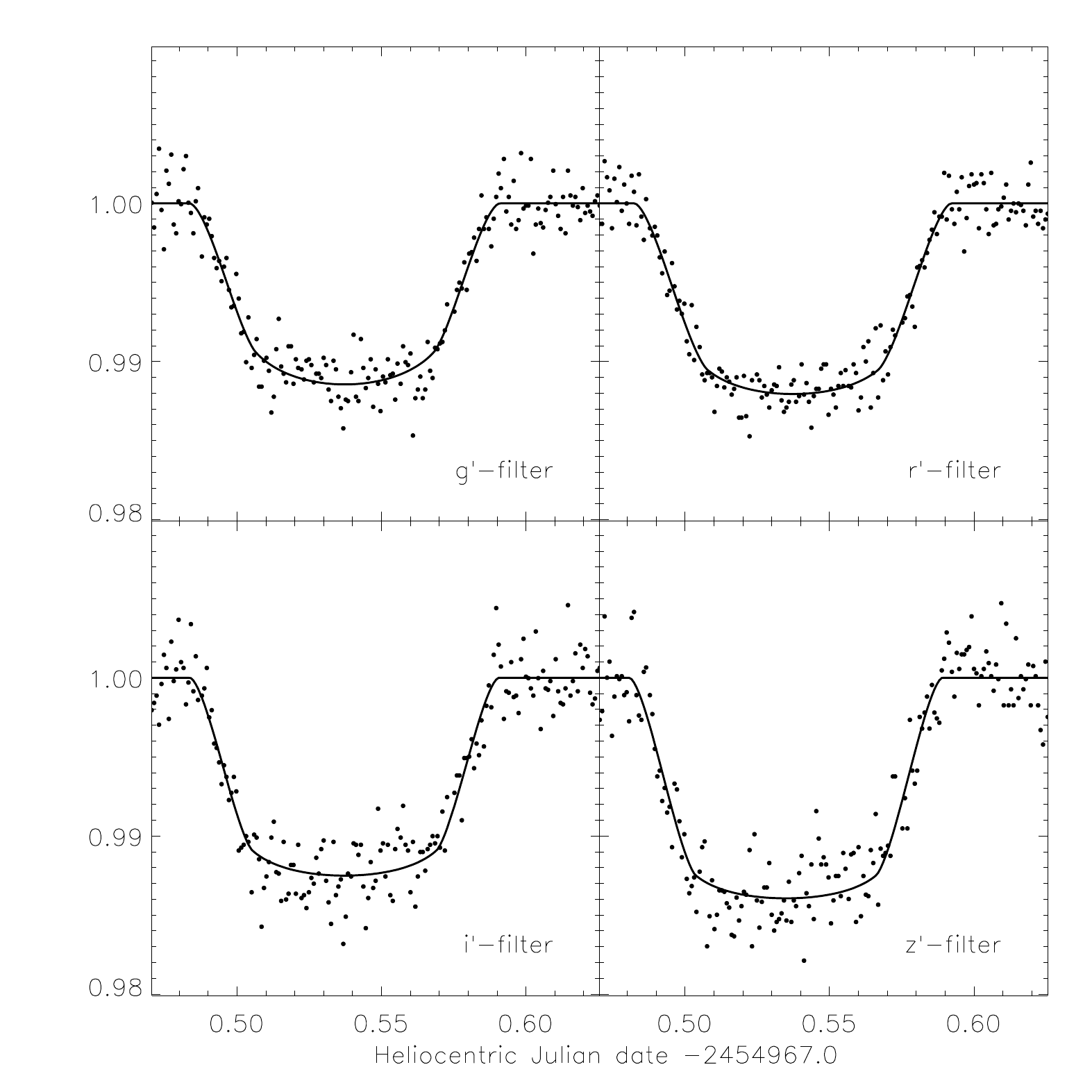}
\label{fig:subfig5}
}
\caption[]{\label{supermegaplot}All light curves in the four optical channels. The data were recorded during the nights of January 28, 2008 (a) 
(by \citealt{Snellen09}), April 10 (b), April 15 (c), 
April 20 (d), April 25 (e) 2009 and May 15 (f) 2009. The continuous lines represent the best fit to the respective data.}
\end{figure*}

For comparison, we also performed aperture photometry using \emph{mupipe} but found that the 
light curves produced with IRAF show a smaller scatter. Therefore we only used the light curves produced
with IRAF for the analysis. Comparing the light curves in the four optical channels, we found that
 the r' filter shows the best accuracy, with a photometric accuracy 
of \mbox{1.5 mmag}, while we achieved \mbox{1.8 mmag} for the g' channel, and \mbox{2.0 mmag} for the i' and z' channels. 
We present all light curves in Figure \ref{supermegaplot}.

\section{Determination of System Parameters}
\label{sec:detsys}
\begin{figure}
\begin{center}
   \includegraphics [width=\linewidth] {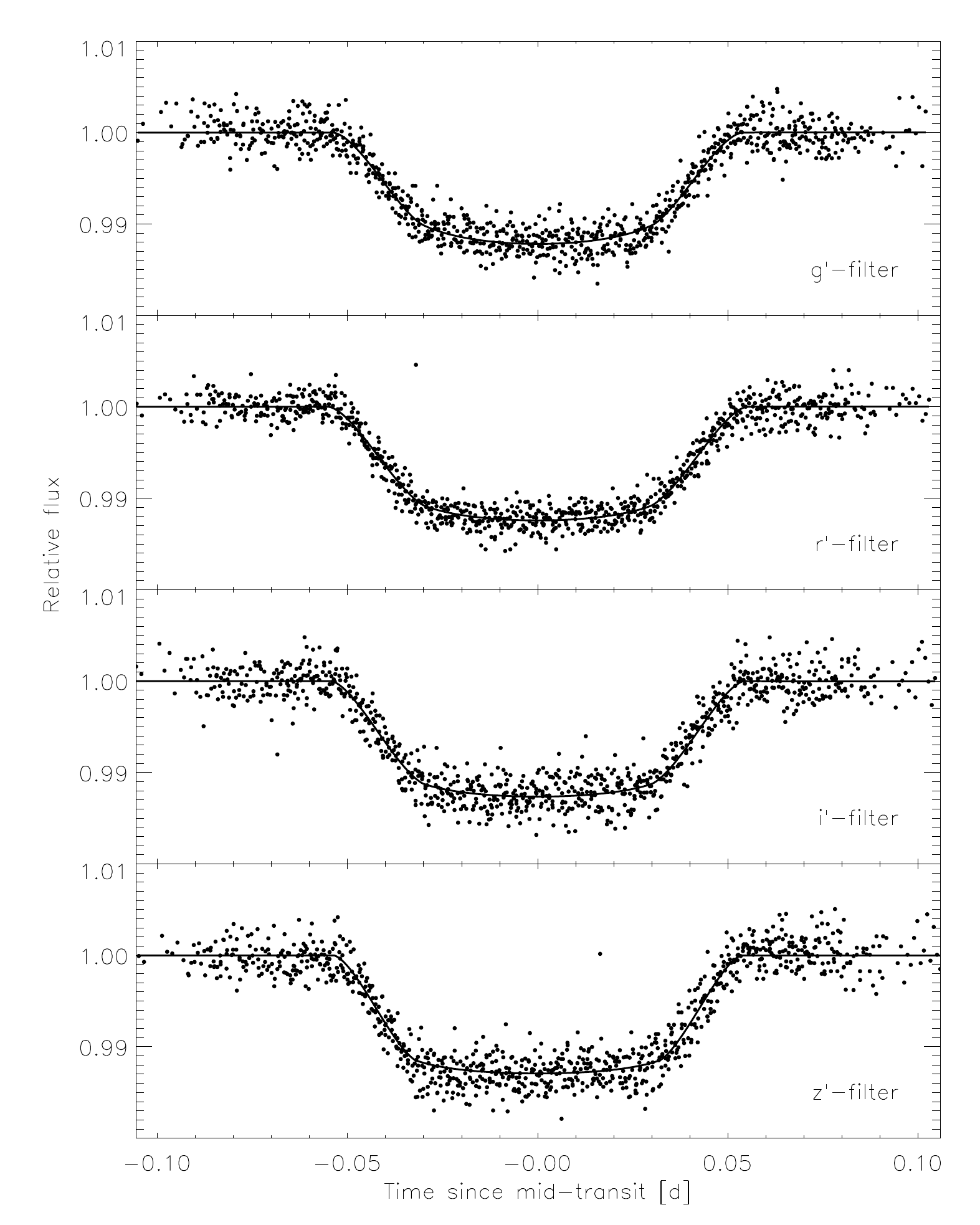} 
   \caption{\label{fig:phazeplot}The phased light-curves of OGLE2-TR-L9. From top to bottom, the filters are: \emph{g'}, \emph{r}', \emph{i'}, and \emph{z'}.
	    The phased light curves also contain the new reduction of the data from Snellen (2009). The best fit models are
	    depicted as continuous lines.}
\end{center}
\end{figure}

The light curve fits were performed using the model for a planetary transit given by \citet{Mandagol}. 
We assumed a quadratic limb darkening law and chose the coefficients according to \citet{Claret04} for a star 
with metallicity $[Fe/H]=0.0$, surface gravity $log(g)=4.5$ and effective temperature $T_{eff}=7000$ $K$. 
We found the best fit to our data by minimizing the $\chi^2$ using a downhill simplex algorithm as implemented 
in the AMOEBA code \citep{Press92}. Our goal was to find the best-fitting values for the central transit time $t_{c}$,  
the ratio of the planetary and stellar radii $r_{p}/r_{\ast}$, the planetary semi-major axis in units of the stellar
radius $a/r_{\ast}$, and the orbital inclination $inc$. The above parameters have the advantage
that they can be derived directly from the light curve without assumptions regarding stellar properties other than limb darkening. 
As starting values, we assumed the central transit times calculated according to the ephemeris given by \citet{Snellen09},
and typical values of planetary transits for the other parameters: $r_{p}/r_{\ast}=0.1$, $a/r_{\ast}=20$ and $inc=90^{\circ}$. 
After the best parameters were found for each light curve, we used the resulting mid-transit times to phase fold the data 
producing the light curves shown in Figure \ref{fig:phazeplot}. To improve on the parameters derived from the data in each of
the four filters, these phase folded light curves were again fitted for $r_{p}/r_{\ast}$, $a/r_{\ast}$ and $inc$. 
The final values were found by a combined fit of all available light curves. The 
resulting parameters are shown together with the results for the phased light curves in Table \ref{filterres} and the corresponding model is 
depicted as a continuous line in Figure \ref{fig:phazeplot}.

To estimate the errors of the determined parameters, we used the Bootstrap Monte-Carlo Method \citep{Press92} which works by creating a large number of representations of the 
data by randomly choosing, with replacement, subsets from the original data set. This means that each newly created data set 
contains the same number of points as the original data set but with some points left out and some points duplicated. 
For each bootstrap data set, the best fitting parameters are found using a procedure identical to the analysis
of original data and the errors are calculated 
from the distribution of the results. To get a second estimate for the errors in the central transit time, we kept the model parameters
fixed at their best values and let $t_{c}$ vary until the deviation in the $\chi^2$ exceeded $\Delta\chi^2 = 1$. The resulting errors 
are in good agreement with the errors derived from the Bootstrap Monte-Carlo Method and range from 29 seconds for transits observed 
with a lower cadence to 16 seconds for transits observed with higher cadence. Here, it should be pointed out that the above error treatment
does not incorporate the contribution of correlated low frequency noise \citep{Pont06}. Thus, the errors on the mid-transit points 
are likely to be underestimated. 

\section{Results}
\label{sec:res}
\subsection{Planetary Parameters}
\label{sec:planpar}
\begin{figure}
\begin{center}
   \includegraphics [width=\linewidth] {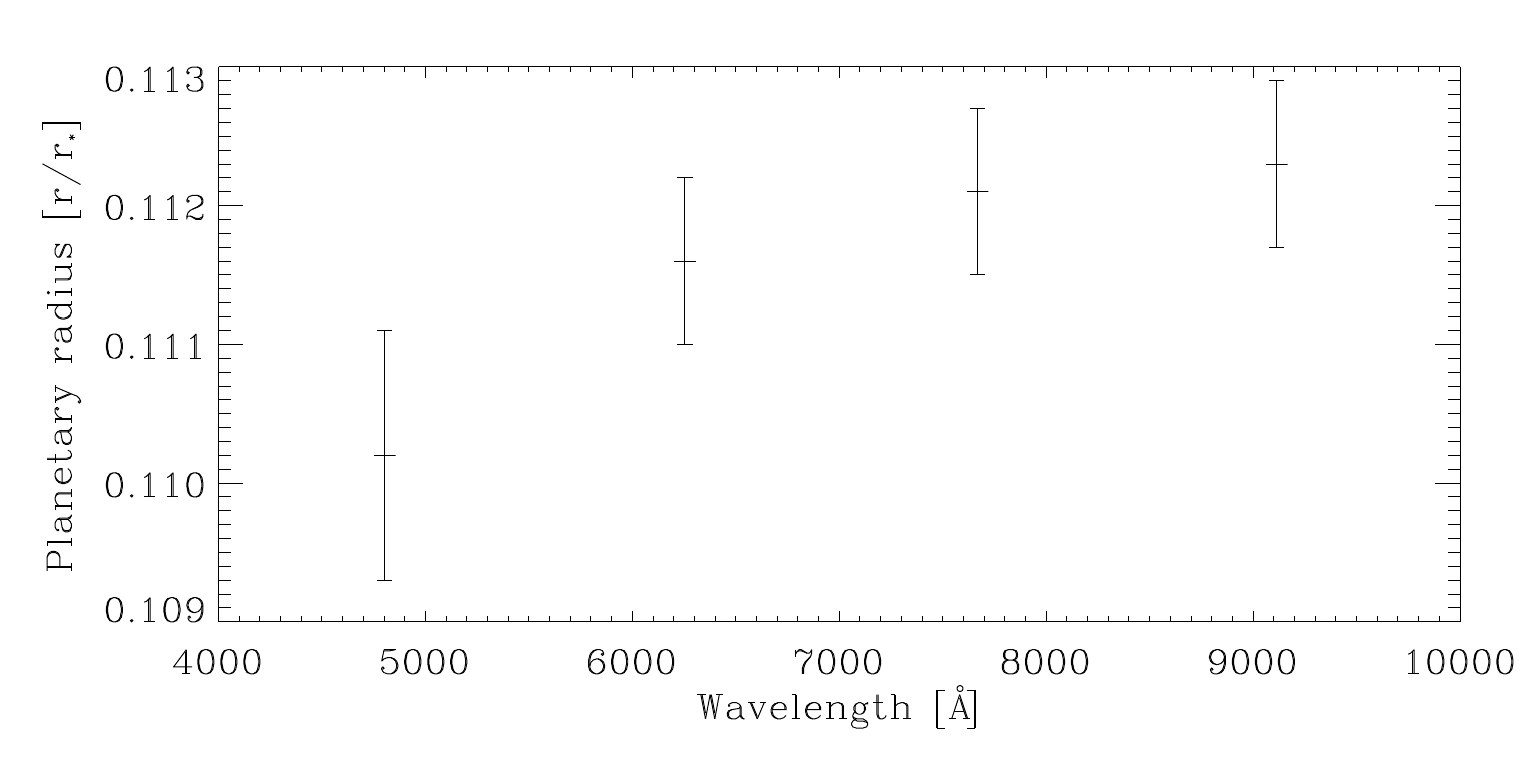} 
   \caption{\label{fig:lamdap}The results for the radius ratio $r_{p}/r_{\ast}$ obtained from the phased light curves 
against the filter wavelengths. The values show a slight trend towards smaller planetary radii with decreasing wavelength.}
\end{center}
\end{figure}

\begin{figure}
\begin{center}
   \includegraphics [width=\linewidth] {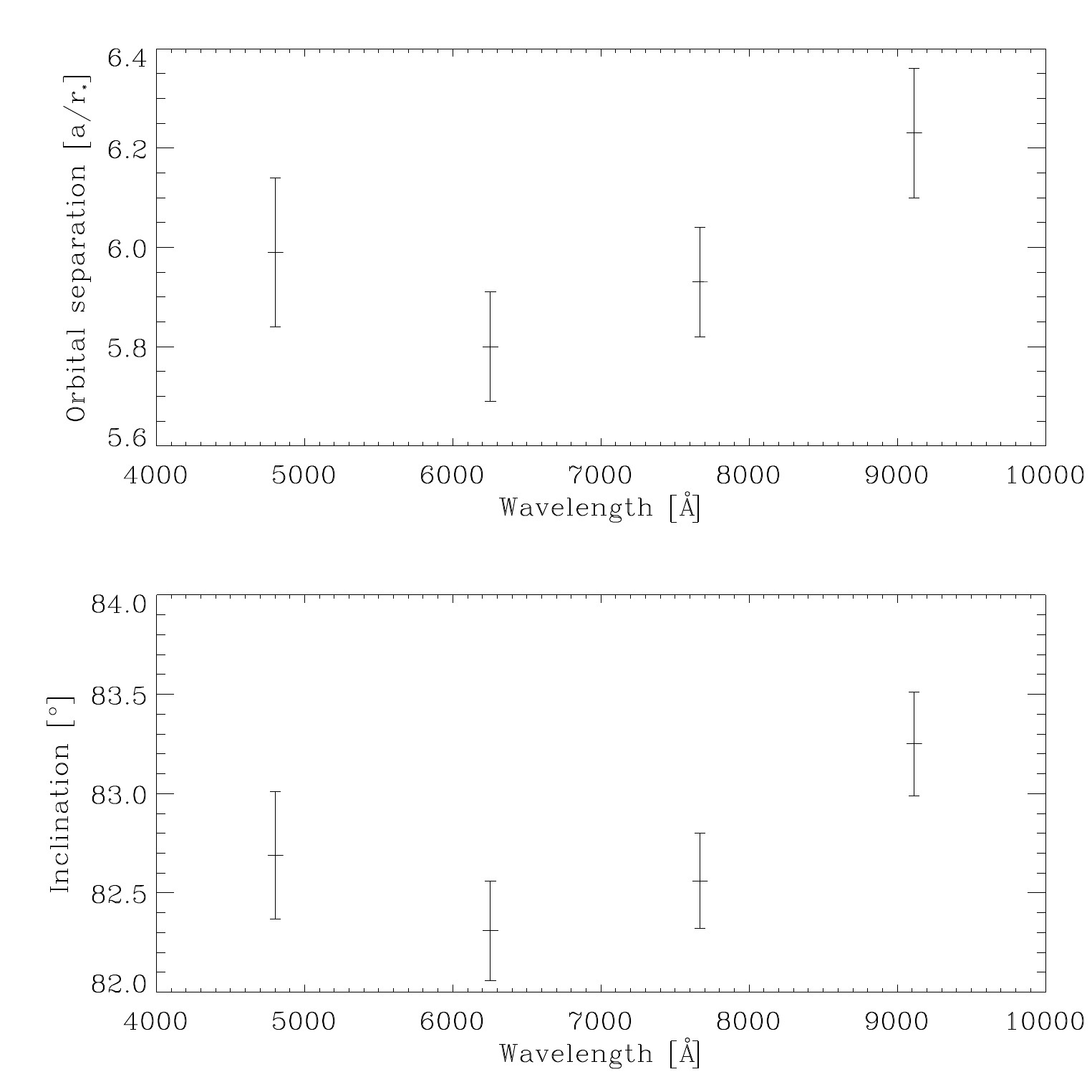} 
   \caption{\label{fig:lamdaai}The results for the planetary semi-major axis (upper panel) and the inclination (lower panel) 
against the filter wavelengths. The fact that the shapes of the variations are very similar shows
 the degeneracy present between the semi-major axis and the inclination: larger solutions for the planetary separation
favor higher inclinations.}
\end{center}
\end{figure}

   \begin{table}
         $$ \begin{array}{p{0.11\linewidth}p{0.15\linewidth}p{0.15\linewidth}p{0.15\linewidth}p{0.15\linewidth}p{0.15\linewidth}}
            \hline
            \noalign{\smallskip}
            Filter:  &  g'  &  r'  & i'  & z' & all \\
            \noalign{\smallskip}
            \hline
            \noalign{\smallskip}
            $r_{p}/r_{\ast} $ & $0.11021$ & $0.1116$ & $0.1121$ & $0.1123$ & $0.1120$\\
            & $\pm0.0009$ & $\pm0.0006$ & $\pm0.0006$ & $\pm0.0006$ & $0.003$ \\
            $a/r_{\ast}$ & $5.99$ & $5.80 $ & $5.93$ & $6.23$ & $5.88$\\
            & $\pm0.15$ & $\pm0.11$ & $\pm0.11$ & $\pm0.13$ & $\pm0.06$\\
            inc [$^\circ$]& $82.69$ & $82.31$ & $82.56$ & $83.25$ & $82.47$ \\
            & $\pm0.32$ & $\pm0.25$ & $\pm0.24$ & $\pm0.26$ & $\pm0.12$ \\
            \noalign{\smallskip}
            \hline
         \end{array}
     $$ 
    \caption[]{ \label{filterres}The results for the planetary to stellar radius ratio, the semi-major axis in units of the stellar radius, and the 
inclination for the filters used next to the final parameters produced by a combined fit of all available light curves.}
   \end{table}

   \begin{table}
         
        $$ \begin{array}{p{0.45\linewidth}p{0.45\linewidth}}
            \hline
            \noalign{\smallskip}
	     Parameter & \\
            \noalign{\smallskip}
            \hline
            \noalign{\smallskip}
	    Semi-major axis [AU] &  $0.0418\pm0.0015$\\    
	    Planetary radius [$R_{jup}$] & $1.67\pm0.05$ \\
	    Inclination [$^\circ$] & $82.47\pm0.12$ \\
	    Transit duration [d] & $ 0.1092\pm0.0021 $ \\
	    Period [d] & $2.48553417\pm 6.4 \times 10^{-7}$ \\
            $T_{0}$ [HJD] & $2454492.80008\pm0.00014$ \\
	    Stellar density [$\rho_{\odot}$] & $0.442\pm0.014$ \\
            \noalign{\smallskip}
            \hline
         \end{array}
     $$ 
    \caption[]{\label{tab:final}The final results for the planetary parameters derived from the data presented in this work.}
\end{table}

In Table \ref{filterres}, we present the results for the phased g', r', i', and z' 
light curves. As can be seen in Figure \ref{fig:lamdap}, the values for the planetary radius mainly agree within one sigma,
 apart from the value derived from the g' filter which lies slightly below.
While values for the planetary semi-major axis and inclination found with the g', r' and i' photometry agree within one sigma, 
the values derived from the z' band observations favor a larger orbital separation combined with a higher inclination (see Figure \ref{fig:lamdaai}). 

As pointed out by \citet{Seager03}, the mean stellar density can be derived directly from the light curve shape and the planetary period.
Following the description of \citet{Seager03}, we find the stellar density to be $\rho_{\ast}=0.442\pm0.014$ $[\rho_{\odot}]$, which is in good agreement
with the previously published value.

To find the planetary parameters $a_{p}$ and $r_{p}$, we adopted the stellar parameters from \citet{Snellen09} 
and combined them with the best-fit values obtained from our analysis. The results and their respective errors can be found 
in Table \ref{tab:final}. All values are in good agreement with the values published in \citet{Snellen09} except for $a_{p}$ 
and $inc$ where our new results are significantly different. The reason for this discrepancy is an error in the calculation
of $a_{p}$ in \citet{Snellen09} where $1/M_{\ast}$ was used instead of $M_{\ast}$ when transforming the measured period 
into a semi-major axis using Kepler's third Law. The error in semi-major axis propagated into a wrong inclination value. 
We recommend to use the values published here for future work because of this calculation error in \citet{Snellen09} 
and because the new values are based on a significantly larger dataset.

\subsection{A New Ephemeris}
\label{sec:eph}

To derive a new ephemeris for OGLE2-TR-L9 b, we used all six observed transits of OGLE2-TR-L9 together with 
an ephemeris derived from the OGLE-II dataset by \citet{Snellen07}. It should be noted that this ''mid-transit point`` 
is derived from an OGLE-II light curve which was constructed from $\sim 500$ data points with irregular cadence collected 
over a period of 3.5 years. From our new analysis of the transit observed by \citet{Snellen09}  on January 27, 2008,
we find the central transit time to be HJD $2454492.80086\pm0.00033$ instead of HJD $2454492.79765\pm0.00039$.
All mid-transit times known for OGLE2-TR-L9 can be found in Table \ref{tab:tcs}. Fitting a constant period to all 
available mid-transit times gives the following new ephemeris: 

\begin{eqnarray*}
T_{c}(E) [HJD]=2454492.80008(\pm0.00014) \\+ 2.48553417(\pm 6.4 \times 10^{-7}) E.
\end{eqnarray*}

\begin{figure}
\centering
   \includegraphics [width=\linewidth]{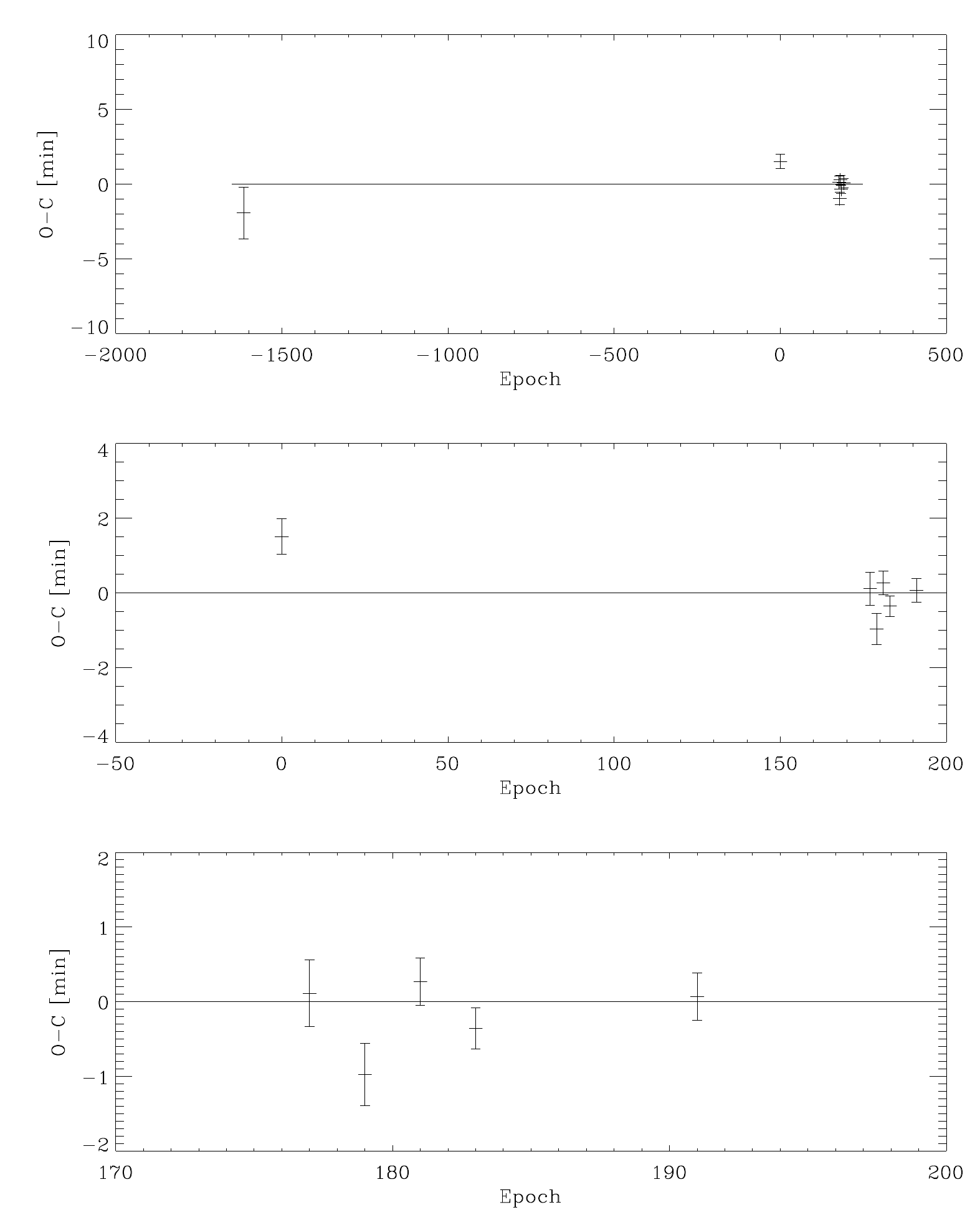} 
   \caption{\label{fig:TTVall}The O-C diagram for the ephemeris calculated from all known mid-transit times of OGLE2-TR-L9 b. In the 
upper panel, all points are shown, while the lower two panels zoom in on the new points together with the point found from the transit
in 2008 (middle panel) and solely the new points (lower panel). The mid-transit points of the epochs 0 and 179 occur 1.5 min later and 1.0 min
earlier than expected, respectively.}
\end{figure}

\begin{figure}
\centering
   \includegraphics [width=\linewidth]{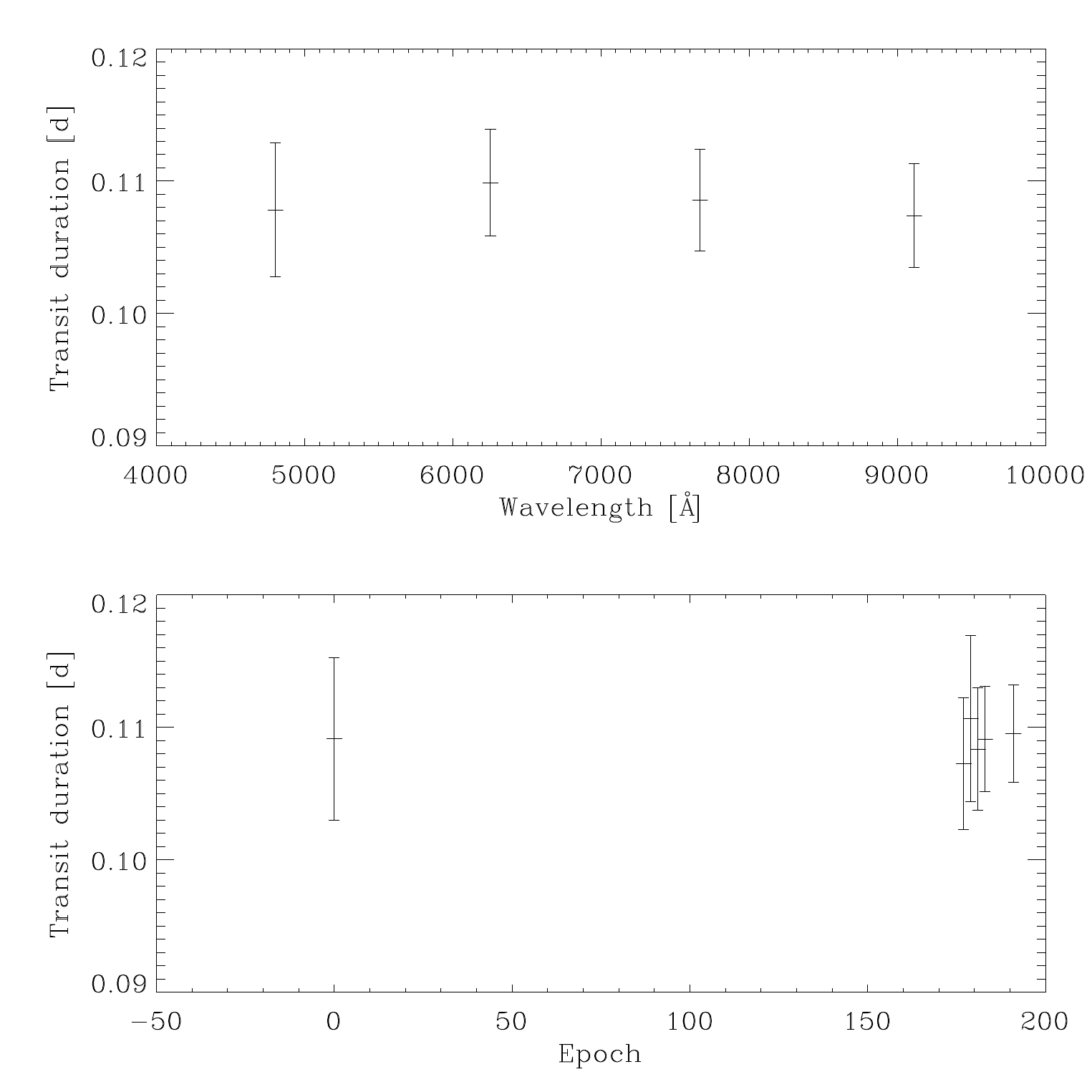} 
   \caption{\label{TDVplot}The transit durations obtained from the phased light curves in the four filters (upper panel) and the 
light curves for each transit event (lower panel). All values agree within their errors.}
\end{figure}

   \begin{table}
        $$ \begin{array}{p{0.35\linewidth}p{0.35\linewidth}}
            \hline
            \noalign{\smallskip}
            $t_{c}$ [HJD - 2400000] & Reference\\
            \noalign{\smallskip}
            \hline
            \noalign{\smallskip}
	    $50478.661\pm0.0012$ &  \citet{Snellen07} \\
	    $54492.80086\pm0.00033$ & this work \\
	    $54932.73944\pm0.00031$ & this work \\
	    $54937.70976\pm0.00029$ & this work \\
	    $54942.68169\pm0.00022$ & this work \\
	    $54947.65232\pm0.00019$ & this work \\
	    $54967.53689\pm0.00022$ & this work \\
            \noalign{\smallskip}
            \hline
         \end{array}
     $$ 
    \caption{\label{tab:tcs}All mid-transit times of OGLE2-TR-L9 known up to date.}
\end{table}

\subsection{Transit Timing and Transit Duration Variations }

Finally, it is possible to search for TTVs. To do so, the observed minus calculated (O-C)
values with reference to the ephemeris determined above are computed for each of the given mid-transit points. 
The result is shown in Figure \ref{fig:TTVall}. Here epoch -1615 corresponds to the ephemeris derived by \citet{Snellen07}
from the OGLE-II data, epoch 0 corresponds to the corrected mid-transit time from \citet{Snellen09} and the last five epochs 
correspond to the newly observed transits. It can be seen that the transit times show some deviations from a constant period 
as the transit of epoch 0 occurs 1.5 min later and the transit of epoch 179 occurs 1.0 min earlier than expected. Fitting a 
constant period to the data as done in Section \ref{sec:eph} gives $\chi^2_{red} = 3.86$ yielding a confidence level of $> 99.75 \%$. 
However, one must take into account that the errors on the central transit times are likely to be underestimated due to correlated 
noise (see Section \ref{sec:detsys}).

Next to variations in the transit timing, we also searched our data for variations in the transit duration which could hint at 
the existence of exo-moons \citep{Kipping09}. For this purpose,
we measured the transit durations based on the best models of both the phased light curves and the light curves derived from 
each transit event. The results are illustrated in Figure \ref{TDVplot} and show no detectable variation.

\section{Conclusion}

We have gathered five full transits of OGLE2-TR-L9 b with the GROND multichannel imager mounted on the ESO/MPG 2.2 m telescope at La Silla Observatory. 
With this new data, and including the transit observed by \citet{Snellen09} in our analysis, we recalculated the parameters of OGLE2-TR-L9 b and found 
different semi-major axis and inclination values compared to \citet{Snellen09}. This is due to a calculation error in the 
previous work and thus the values presented in this work should be used in the future.
We studied the central transit times and transit durations for any variations that could be attributed to a perturbing body in the 
OGLE2-TR-L9 system. While the transit durations agree for all transits, the mid-transit points show indications of period
variations.

\begin{acknowledgements}
We thank P. Afonso, F. Olivares and A. Rossi from the GROND team for their valuable support during the observations. \newline
Based on observations collected at the European Organisation for Astronomical Research in the Southern Hemisphere, Chile
(083.A-9005(A)) during MPG guaranteed time. \newline Part of the funding for GROND (both hardware as well as personnel) was generously 
granted from the Leibniz-Prize to Prof. G. Hasinger (DFG grant HA 1850/28-1). 

\end{acknowledgements}

\bibliographystyle{aa}
\bibliography{bibl}

\end{document}